\documentclass[journal=nalefd,manuscript=article,layout=twocolumn]{achemso}

\usepackage[T1]{fontenc}
\usepackage{chemformula}
\usepackage{graphicx}
\usepackage{amsmath}
\usepackage{float}
\usepackage{layouts}
\usepackage{hyperref}
\usepackage{bm}
\makeatletter
\makeatletter
\renewcommand*\acs@contact@details{%
  { \sffamily *E-mail: \acs@email@list }%
  \acs@number@list
}
\makeatother

\title{Spin-State Selective Excitation in Spin Defects of Hexagonal Boron Nitride}

\let\oldmaketitle\maketitle
\let\maketitle\relax

\author{Mohammad Abdullah Sadi}
\email{msadi@purdue.edu}
\affiliation[Purdue University]{Elmore Family School of Electrical and Computer Engineering, Purdue University, West Lafayette, Indiana 47907, USA}
\alsoaffiliation[Purdue University]{Department of Physics and Astronomy, Purdue University, West Lafayette, Indiana 47907, USA}

\author{Luca Basso}
\affiliation[Sandia National Laboratories]{Center for Integrated Nanotechnologies, Sandia National Laboratories, Albuquerque, New Mexico 87123, USA}

\author{David A Fehr}
\affiliation[University of Iowa]{Department of Physics and Astronomy, University of Iowa, Iowa City, Iowa 52242, USA}

\author{Xingyu Gao}
\affiliation[Purdue University]{Department of Physics and Astronomy, Purdue University, West Lafayette, Indiana 47907, USA}

\author{Sumukh Vaidya}
\affiliation[Purdue University]{Elmore Family School of Electrical and Computer Engineering, Purdue University, West Lafayette, Indiana 47907, USA}

\author{Emmeline G Riendeau}
\affiliation[Sandia National Laboratories]{Center for Integrated Nanotechnologies, Sandia National Laboratories, Albuquerque, New Mexico 87123, USA}
\alsoaffiliation[University of Chicago]{Department of Physics, University of Chicago, Chicago, Illinois 60637, USA}

\author{Gajadhar Joshi}
\affiliation[Sandia National Laboratories]{Center for Integrated Nanotechnologies, Sandia National Laboratories, Albuquerque, New Mexico 87123, USA}

\author{Tongcang Li}
\affiliation[Purdue University]{Elmore Family School of Electrical and Computer Engineering, Purdue University, West Lafayette, Indiana 47907, USA}
\alsoaffiliation[Purdue University]{Department of Physics and Astronomy, Purdue University, West Lafayette, Indiana 47907, USA}
\alsoaffiliation[Purdue University]{Birck Nanotechnology Center, Purdue University, West Lafayette, Indiana 47907, USA}
\alsoaffiliation[Purdue University]{Purdue Quantum Science and Engineering Institute, Purdue University, West Lafayette, Indiana 47907, USA}

\author{Michael E Flatté}
\affiliation[University of Iowa]{Department of Physics and Astronomy, University of Iowa, Iowa City, Iowa 52242, USA}

\author{Andrew M Mounce}
\email{amounce@sandia.gov}
\affiliation[Sandia National Laboratories]{Center for Integrated Nanotechnologies, Sandia National Laboratories, Albuquerque, New Mexico 87123, USA}

\author{Yong P Chen}
\email{yongchen@purdue.edu}
\affiliation[Purdue University]{Elmore Family School of Electrical and Computer Engineering, Purdue University, West Lafayette, Indiana 47907, USA}
\alsoaffiliation[Purdue University]{Department of Physics and Astronomy, Purdue University, West Lafayette, Indiana 47907, USA}
\alsoaffiliation[Purdue University]{Birck Nanotechnology Center, Purdue University, West Lafayette, Indiana 47907, USA}
\alsoaffiliation[Purdue University]{Purdue Quantum Science and Engineering Institute, Purdue University, West Lafayette, Indiana 47907, USA}

\keywords{Quantum sensing, hexagonal boron nitride, spin defects, microwave polarization, spin-state control}

\begin{document}
\twocolumn[
\begin{@twocolumnfalse}
\oldmaketitle
\begin{abstract}
Hexagonal boron nitride (hBN) has emerged as a promising two-dimensional platform for quantum sensing, due to its optically addressable spin defects, such as the negatively charged boron vacancy (V$_\text{B}^-$). Despite hBN being transferrable to close proximity to samples, spectral overlap of spin transitions due to large hyperfine interactions has limited its magnetic sensitivity. Here, we demonstrate spin-selective excitation of V$_\text{B}^-$ spin defects in hBN driven by circularly polarized microwave. Using a cross-shaped microwave resonance waveguide, we superimpose two orthogonally linearly polarized microwave shifted in phase from a RFSoC FPGA to generate circularly polarized microwaves. This enables selective spin $|0\rangle\rightarrow|-1\rangle$ or $|0\rangle\rightarrow|1\rangle$ excitation of V$_\text{B}^-$ defects, as confirmed by optically detected magnetic resonance experimentally and supported computationally. We also investigate the influence of magnetic field on spin-state selectivity. Our technique enhances the potential of hBN platform for quantum sensing through better spin state control and magnetic sensitivity particularly at low fields.
\end{abstract}

\end{@twocolumnfalse}
\vspace{2em}
]

\begin{figure*}[t]
\centering
\includegraphics[width=0.98\textwidth]{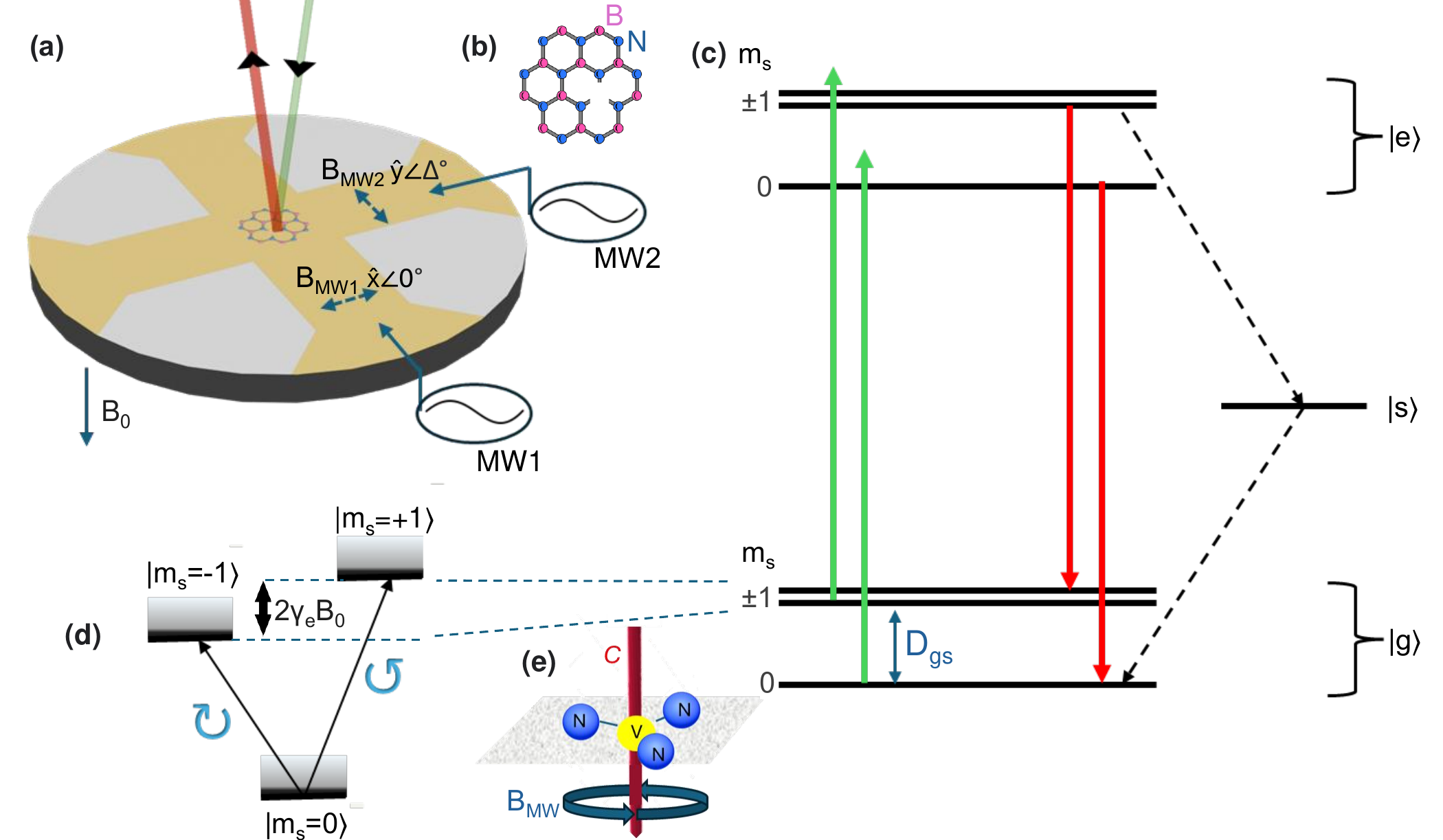}
\caption{Experimental configuration, energy levels and mechanism of spin-state selective excitation in V$_\text{B}^-$ defects in hexagonal boron nitride (hBN). (a) Gold on sapphire device generating circularly polarized microwaves through superposition of orthogonal linearly polarized fields. Center: multilayer hBN containing spin defects. (b) Lattice structure of hBN with V$_\text{B}^-$ . (c) Electronic energy level diagram of V$_\text{B}^-$ showing S=1 triplet states and S=0 metastable state mediating ODMR (green: optical excitation, red: radiative decay). $|e\rangle$, $|s\rangle$ and $|g\rangle$ represent excited, metastable, and ground states. (d) Ground state energy levels showing selective spin-state excitation via angular momentum transfer from circularly polarized microwaves. (e) The boron vacancy is symmetric along c-axis of hBN, necessitating orthogonal circularly polarized microwaves as generated by the device in (a) to drive spin-selective transition in (d).}

\label{fig:figure1}
\end{figure*}

Quantum sensing technologies have seen remarkable advancements in recent years, with solid-state spin defects emerging as particularly promising platforms \cite{pirandola2018advances,schirhagl2014nitrogen,degen2017quantum,budker2007optical}. While nitrogen-vacancy (NV) centers in diamond have been widely studied, their applications are limited by challenges in fabrication, integration, and achieving close proximity to samples of interest \cite{romach2015spectroscopy,sangtawesin2019origins}.

In this context, two-dimensional van der Waals materials, particularly hexagonal boron nitride (hBN), have garnered significant attention \cite{azzam2021prospects,ren2019review,caldwell2019photonics,kubanek2022coherent}. hBN's layered structure allows for exfoliation and direct transfer onto a wide variety of samples, enabling better proximity for quantum sensing. This capability, combined with hBN's wide bandgap and exceptional stability, positions it as a compelling platform for quantum sensing and information applications \cite{cassabois2016hexagonal,kianinia2017robust,jungwirth2016temperature,xue2018anomalous}.
The negatively charged boron vacancy (V$_\text{B}^-$) in hBN, an optically addressable spin defect, was first identified in 2020 \cite{gottscholl2020initialization}. Subsequent theoretical studies \cite{reimers2020photoluminescence,chen2021photophysical,ivady2020ab} and optically detected magnetic resonance (ODMR) experiments \cite{gao2021high,yu2022excited,mu2022excited,mathur2022excited} have elucidated its atomic and electronic structure. The V$_\text{B}^-$ defect is formed by a vacant boron site surrounded by three nitrogen atoms in the hBN lattice (Fig.~\ref{fig:figure1}(b)).  As illustrated in the energy level diagram of Fig.~\ref{fig:figure1}(c), this defect possesses a spin-1 ground state and is optically addressable at room temperature, contributing to its suitability for quantum sensing applications.

Despite these promising characteristics, the potential of V$_\text{B}^-$ defects for quantum sensing has been constrained by large hyperfine interactions. These interactions lead to broad ODMR with spectrally overlapped hyperfine spin transitions, especially at low magnetic
fields, posing challenges for achieving high magnetic sensitivity. Overcoming this limitation is crucial for fully leveraging the sensing capabilities and applicability of hBN-based quantum sensors. 

Circularly polarized microwave fields have long been recognized for their use in manipulating spin states \cite{nelson1957circularly}, with recent applications in various quantum systems \cite{henderson2008high,alegre2007polarization,yaroshenko2020circularly}. Applying this technique to V$_\text{B}^-$ defects in hBN presents an opportunity to address the challenges posed by hyperfine broadening and enhance the system's sensing capabilities.

In this study, we explore the use of circularly polarized microwaves for spin-state selective excitation in V$_\text{B}^-$ defects. We generate these defects through irradiation of hBN flakes with 2.5 keV He$^+$ ions. We employ a cross-shaped microwave waveguide and dual microwave signals with controlled phase difference generated from FPGA  operated with QICK-DAWG package\cite{riendeau2023quantum} to create polarization-tunable microwave fields for spin manipulation (Fig. 1(a)).  We also investigate the influence of magnetic field strength on spin-state selectivity. Our work demonstrates spin-state selective excitation in V$_\text{B}^-$ spin defects of hBN, addressing the challenge of spectral overlap caused by large hyperfine interactions. Our approach enables more control over spin state transitions, which is valuable for advancing hBN-based quantum sensing applications. This improved spin-state transition manipulation technique, combined with hBN's inherent advantages as a 2D platform, represents a significant step towards realizing high-resolution quantum sensing, with implications for nanoscale magnetic field detection, quantum information processing, and sensing applications in areas such as material science and biology.

\begin{figure*}[t]
\centering
\includegraphics[width=0.98\textwidth]{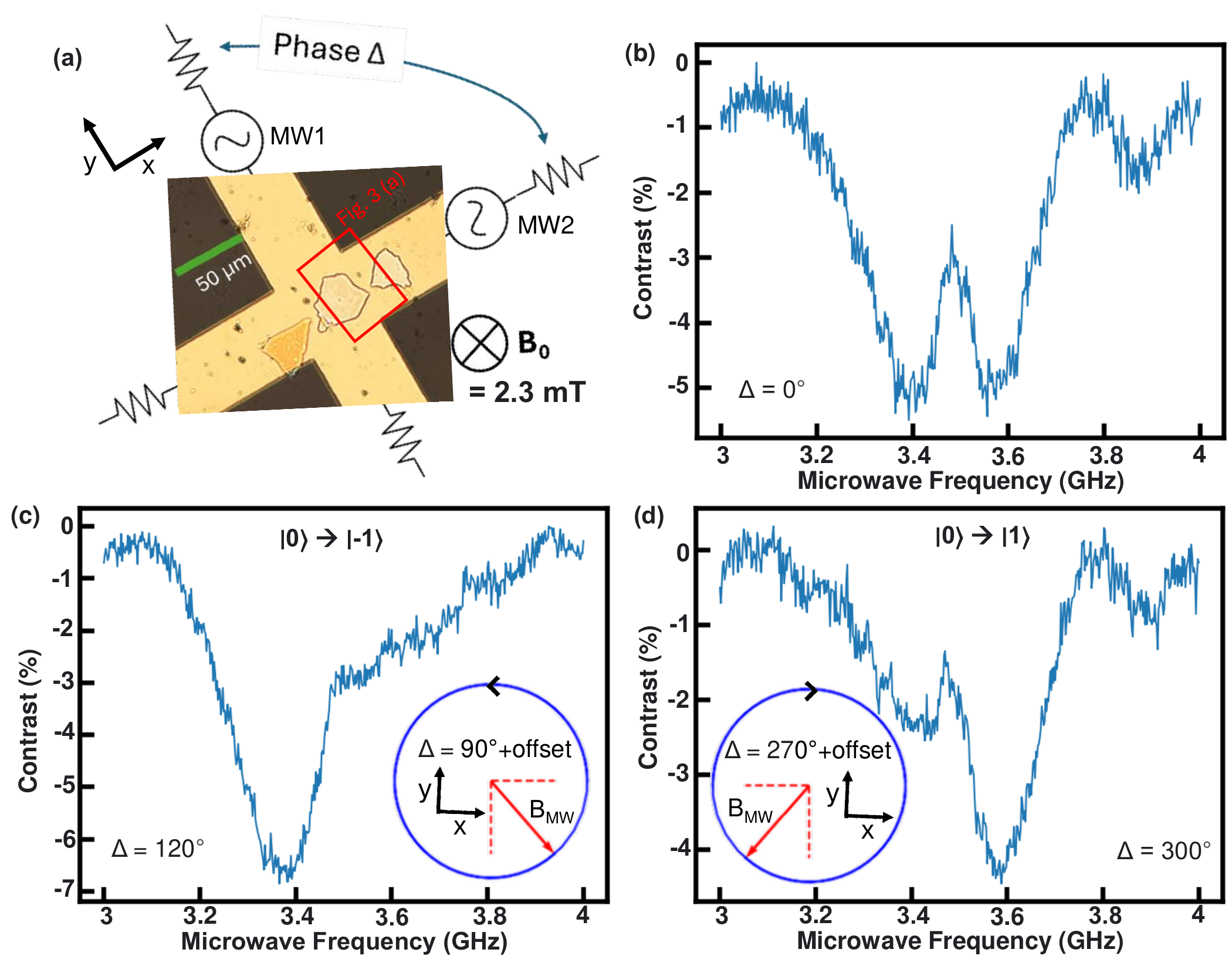}
\caption{ODMR spectroscopy of V$_\text{B}^-$ defects in hBN under controlled microwave polarization. (a) Optical image of orthogonal transmission lines (waveguides) carrying linearly polarized microwaves with controlled phase difference  $\Delta$, with hBN flake at the intersection, under a magnetic field of 2.3 mT applied perpendicular to the surface. (b) ODMR spectrum with $\Delta$ = 0° applied (near linear microwave polarization) (c) ODMR spectrum at $\Delta$ = 120° (effectively 90°, thus counter-clockwise circular polarization) showing selective $|0\rangle$ to $|-1\rangle$ excitation. (d) ODMR spectrum at $\Delta$ = 300° (effectively 270°, thus clockwise circular polarization), showing mostly selective $|0\rangle$ to $|1\rangle$ excitation.}
\label{fig:figure2}
\end{figure*}

The behavior of V$_\text{B}^-$ defects in hBN is governed by the ground-state electron spin Hamiltonian $H_{gs}$, which encompasses terms associated with Zero Field Splitting (ZFS), electron Zeeman splitting, and electron-nuclear spin hyperfine interaction \cite{gottscholl2020initialization,gao2022nuclear}:

\begin{equation}
\begin{split}
H_{gs} = D_{gs} \left[ S_z^2 - \frac{S(S+1)}{3} \right] + E_{gs} \left( S_x^2 - S_y^2 \right) & \\ + \gamma_e \mathbf{B} \cdot \mathbf{S} + \sum_{k=1,2,3} \mathbf{S}\cdot\bar{\bar{A}}_k\cdot\mathbf{I}_k
\end{split}
\end{equation}

Here, $D_{gs}$ ($\approx h \times 3.48 \text{ GHz}$) is the ground state longitudinal ZFS parameter, $E_{gs}$ ($\approx h \times 50 \text{ MHz}$) is the transverse ZFS parameter, $S=1$ is the electron spin quantum number while $S_z$, $S_x$, and $S_y$ are the corresponding spin operator components for a spin-1 system, $\gamma_e$ is the electron-spin gyromagnetic ratio (with Landé $g$-factor $g = 2$, $\gamma_e = g \frac{e}{2m_e\hbar}$, where $e$ is electron charge and $m_e$ is electron mass), and $\mathbf{I}_k$ is the nuclear spin-1 vector operator of the three nearest $^{14}$N nuclei \cite{gottscholl2020initialization} indexed by $\mathbf{k}$. The hyperfine interaction tensor $\bar{\bar{A}}_k$ has a magnitude of approximately $h \times 47 \text{ MHz}$ \cite{gottscholl2020initialization,ivady2020ab,gracheva2023symmetry}.

When an external magnetic field $B_0$ is applied perpendicular to the hBN plane, the resonant frequencies for transitions between the $m_s = 0$ and $m_s = \pm1$ states are given by:
\begin{equation}
f_{\pm} = D_{gs}/h \pm \sqrt{E_{gs}^2 + (\gamma_e B_0)^2}/h
\end{equation}
As illustrated in Figure 1(c), circularly polarized microwave radiation can selectively drive either the $m_s = 0 \to m_s = 1$ or $m_s = 0 \to m_s = -1$ transition, depending on its handedness. This selectivity arises from angular momentum conservation rules. 

\begin{figure*}[t]
\centering
\includegraphics[width=0.98\textwidth]{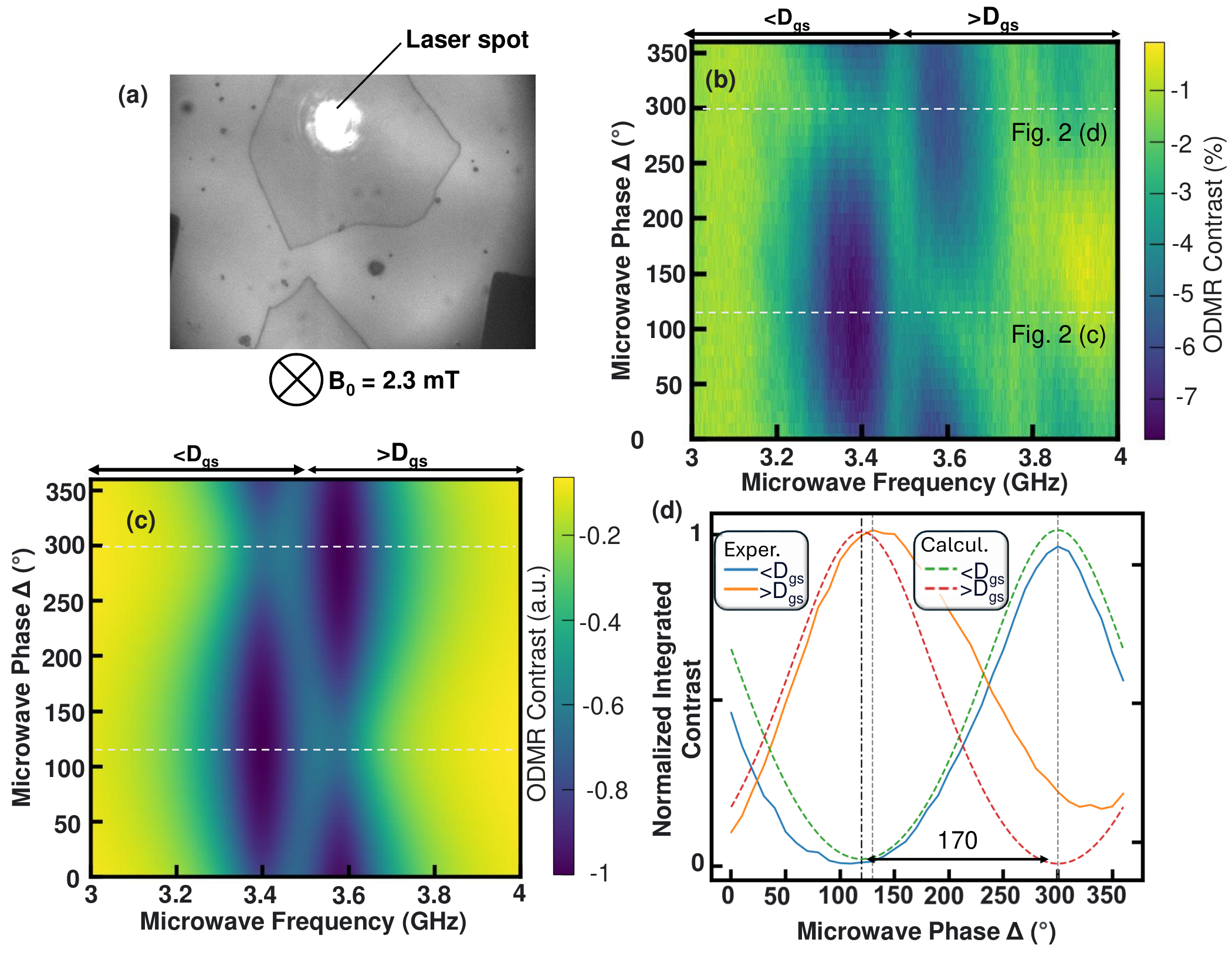}
\caption{Continuous microwave phase-dependent ODMR modulation of V$_\text{B}^-$ defects in hBN. (a) Optical image showing the laser spot for ODMR measurement. (b) Color plot of ODMR spectra evolution versus phase difference ($\Delta^\circ$) between orthogonal linearly polarized microwaves (c) Corresponding Lindblad calculations of the spectral evolution. White dashed lines in both (b) and (c) mark maximum spin selectivities. (d) Normalized integrated contrast for frequencies below  $D_{gs}$ ($|0\rangle\rightarrow|-1\rangle$) and above $D_{gs}$ ($|0\rangle\rightarrow|1\rangle$) with solid (dashed) lines depicting experimental (calculation) results.}

\label{fig:figure3}
\end{figure*}

To achieve this selective excitation, we implemented a cross-shaped microwave waveguide that generates circularly polarized microwaves. At the center of the cross, the magnetic field from the microwave is described by:
\begin{equation}
\vec{\text{B}}_\text{MW} = \text{B}_\text{MW1} \, \hat{x} \sin(\omega t) + \text{B}_\text{MW2} \, \hat{y} \sin(\omega t + \Delta)
\end{equation}
where $\Delta$ is the phase difference between the two orthogonal components, $\omega = 2\pi f$ is the angular frequency of the microwave (with $f$ being the frequency), $\text{B}_\text{MW1}$ and $\text{B}_\text{MW2}$ are the amplitudes of the magnetic fields from orthogonally linearly-polarized microwaves incoming at the center of the cross, and $t$ is time (see Fig. 1(a)). We ensured approximately equal amplitudes between the orthogonal magnetic fields $\text{B}_\text{MW1}$ and $\text{B}_\text{MW2}$ by adjusting power to obtain reasonably equivalent ODMR signal intensities when independently exciting the hBN flake through each arm of the cross waveguide. The equation then describes circular polarization when $\Delta=90^{\circ}$ or $\Delta=270^{\circ}$, linear polarization when $\Delta=0^{\circ}$ or $\Delta=180^{\circ}$, and elliptical polarization for other angles (modulo $360^{\circ}$). The x-y plane is perpendicular to the symmetry axis "c" of the defect shown in Figure 1 (e).

Our experiments employ continuous wave (CW) ODMR measurements, with contrast determined by the normalized relative photoluminescence change when the microwave drive is turned on. We use 532 nm laser excitation and collect emitted light at wavelengths above 630 nm. The complete optical setup details are provided in the supplementary information (SP). The ODMR transition pathway, shown in Figure 1(c), illustrates the reduction of photoluminescence due to metastable state-mediated phonon transitions (non-radiative decays) when optically excited from m$_s$=$\pm1$ compared to that from m$_s$=0 state, whereas the former can be reached from the m$_s$=0 ground state under microwave at resonant frequencies. The cross waveguide design is a modification of previous gapped waveguides \cite{alegre2007polarization}, adapted for the V$_\text{B}^-$ hBN platform. Unlike the previous capacitively coupled design \cite{alegre2007polarization} with sharp resonance, our modified version incorporates a directly coupled broadband waveguide (Figure 1(a) and supplementary information). This adaptation addresses two key challenges: the broad ODMR signal of V$_\text{B}^-$ defects and the difficulty in precisely tuning resonant frequencies.

To demonstrate spin-state selectivity in V$_\text{B}^-$ defects, we transferred an hBN flake containing these defects to the center of our cross-shaped microwave waveguide, as shown in Fig.~\ref{fig:figure2}(a). This configuration allows for the superposition of orthogonal microwave fields with adjustable phase differences. While ideally the superimposed microwaves should have equal strength, the different paths taken by the signals necessitated some adjustment to the input microwave magnitudes. We applied a perpendicular magnetic field of 2.3 mT to the flake using a permanent magnet, and utilized 50 $\Omega$ terminations to minimize microwave reflections. ODMR spectra collected under these conditions are presented after linear background subtraction. All measurements were performed at room temperature under ambient conditions.

Figure 2(b) presents the measured ODMR spectrum with zero applied phase difference ($\Delta = 0^\circ$) between the orthogonal microwave fields. The observed spectrum exhibits two resonance dips of nearly equal magnitude, indicating close to linearly polarized microwave excitation. However, due to differences in the microwave signal path lengths, some undetermined offset between the applied and the actual phase difference may be present. As we varied the phase difference, we observed optimal spin selectivity for the $|0\rangle \to |-1\rangle$ transition at $\Delta = 120^\circ$ (Figure 2(c)) and for the spin $|0\rangle \to |1\rangle$ transition at $\Delta = 300^\circ$ (Figure 2(d)). This $180^\circ$ separation in optimal phase differences aligns with our expectations based on Equation 3 ($90^\circ$ and $270^\circ$), confirming successful implementation of circularly polarized microwave excitation.

For this experiment, the offset phase between the applied and the actual phase difference of the orthogonal microwaves comes out to be -30°. Thus at $\Delta=0^\circ$ the polarization of microwave at the cross is slightly but not significantly deviated from being linear. Interestingly, the spin-state selectivity exhibits an asymmetry not observed in NV centers in diamond. Lorentzian fitting of the ODMR spectra, by which selectivity is defined as discussed in detail later in the discussion of Figure 4, reveals that the best-case scenario for the spin $|0\rangle \to |-1\rangle$ transition achieves 79.5$\pm$1.7\,\% selectivity, while the spin $|0\rangle \to |1\rangle$ transition reaches 67.4$\pm$1.9\,\% optimal selectivity. This asymmetry suggests an intrinsic difference between these transitions in V$_\text{B}^-$ defects, warranting further investigation.
It is worth noting that the spurious peak around 3900 MHz observed in the ODMR data in Figure 2 likely arises from an aliasing artifact from the FPGA sampling process. This artifact does not affect the overall interpretation of our results but highlights the importance of considering potential experimental limitations in high-frequency measurements.

To gain a more comprehensive understanding of the spin-state selective excitation in V$_\text{B}^-$ hBN, we investigate the continuous modulation of the transitions as a function of microwave phase difference $\Delta$. Fig.~\ref{fig:figure3}(a) presents a magnified view of the experimental setup shown in Fig.~\ref{fig:figure2}(a), clearly illustrating the laser spot of diameter of 20 $\mu$m, at the center of the cross, on an hBN flake approximately 50 nm thick. The color plot in Fig.~\ref{fig:figure3}(b) depicts the evolution in ODMR peak intensities as the phase between the orthogonal microwave fields is continuously tuned, with the white dashed lines corresponding to the maximally selective transitions highlighted in Fig.~\ref{fig:figure2}. 

To corroborate the experimental data, simulations of ODMR were completed using the Lindblad formalism \cite{candido2024theory,patel2024room,candido2024interplay,elko2024near}, implemented using a combination of Mathematica software and the QuTiP Python package. Coherent transitions resulting from microwave-induced Rabi driving were modeled with a Hamiltonian of a similar form to Equation 1, and the optical dynamics in Figure 1 were modeled with Lindblad jump operators and associated optical pumping and relaxation rates \cite{whitefield2024magnetic}. In the Hamiltonian, we used $g$ = 2.002, $D_{gs}$ = 3.49 GHz, and $E_{gs}$ = 65 MHz, with parameters extracted from fitting experimental data. The $E_{gs}$ value reflects minor deviation from ideal C3v symmetry of the defects under experimental conditions due to local structural and/or electronic environment asymmetries.  A dephasing Lindblad jump operator was used to model finite temperature and hyperfine broadening effects on the ODMR linewidths to match the measurements, with a dephasing rate $\sim$ 100 $\mu s^{-1}$. In the simulation a 7-level energy diagram was used, reflecting the S=1, S=1, and S=0 spin character of the respective ground, excited, and metastable spin manifolds. Further details on the Lindblad calculations are provided in the supplementary information.
Figure ~\ref{fig:figure3}(c) presents the corresponding Lindblad calculations of the spectral evolution observed in ~\ref{fig:figure3}(b), with the offset phase of -30° incorporated. The calculations are in good agreement with the experimental data, as in both cases the white dashed lines representing maximum spin selectivities are separated by 180°.

By analyzing the data from Figures ~\ref{fig:figure3}(b) and ~\ref{fig:figure3}(c), the ODMR signal above and below the zero-field splitting frequency ($D_{gs}$=3.49 GHz) is separated. The integrated ODMR contrast for these two frequency windows, plotted in Figure ~\ref{fig:figure3}(d), demonstrates continuous modulation of the spin-state selectivity. For both experimental and calculated results, we normalize the data such that the minimum value in each pair of datasets corresponds to zero and the maximum value corresponds to one. The solid lines represent the normalized experimental data, while the dashed lines show the normalized calculated values. The separation between phases corresponding to peaks in normalized integrated contrast for frequencies below $D_{gs}$ (predominantly associated with $|0\rangle \to |-1\rangle$ transition) and above $D_{gs}$ (predominantly associated with $|0\rangle \to |1\rangle$ transition) is approximately $170^\circ$ in experimental data, while it is precisely $180^\circ$ in the calculated results. This separation has good agreement with the expected phase difference of opposite circular polarizations, generated by the microwave fields as described in Equation 3, each of which is expected to result in maximum selectivity for one of the spin-state transition in the spin defects.

\begin{figure}[!htb]
\centering
\includegraphics[width=\columnwidth]{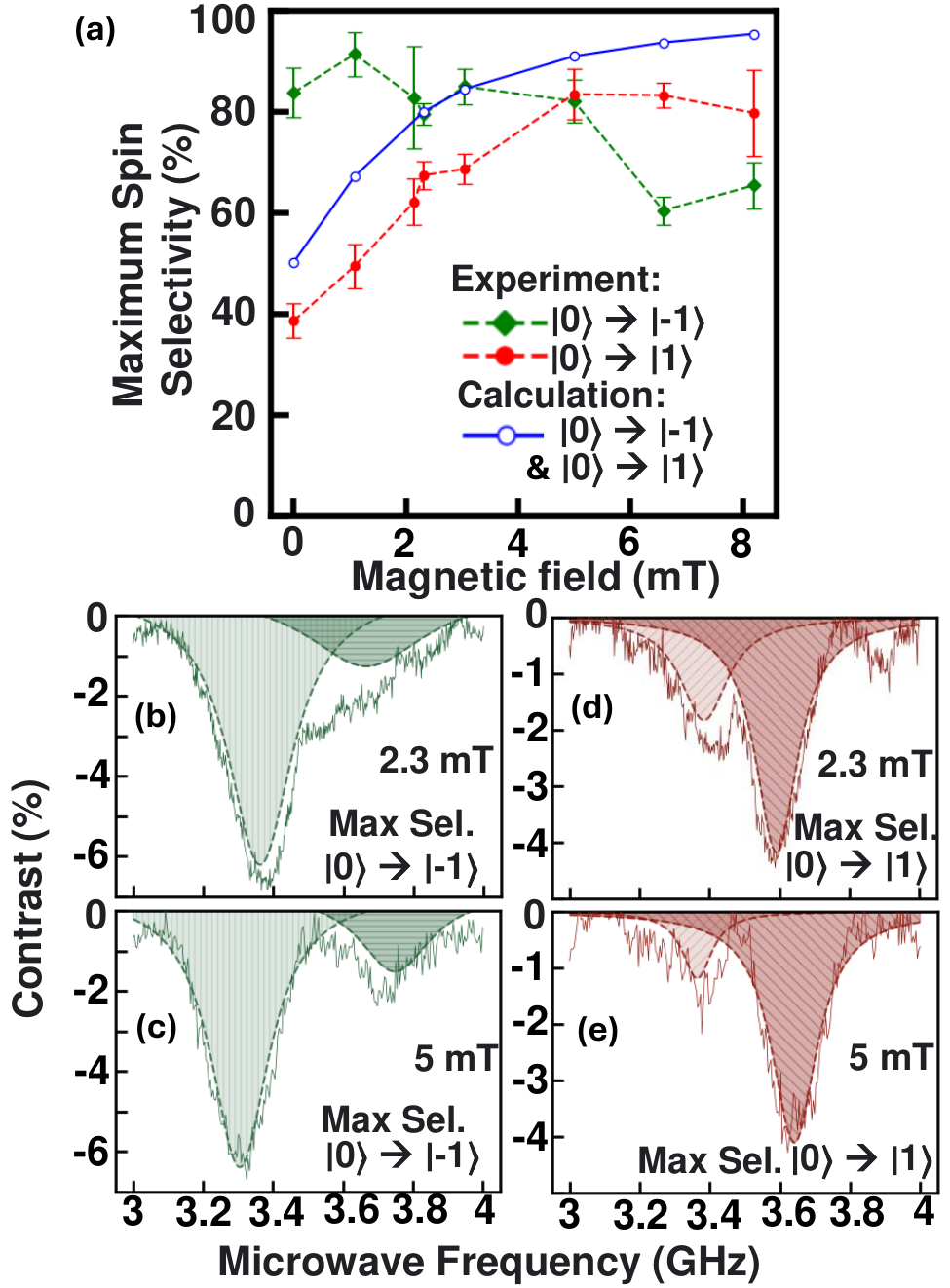}
\caption{Magnetic field dependence of spin-state selectivity in V$_\text{B}^-$ defects in hBN. (a) Maximum spin selectivity versus magnetic field showing experimental data for $|0\rangle\rightarrow|-1\rangle$ (green diamonds) and $|0\rangle\rightarrow|1\rangle$ (red circles) transitions with corresponding calculations (blue circles). (b-e) ODMR spectra at maximum selectivity for $|0\rangle\rightarrow|-1\rangle$ transition of 2.3 mT (b) and 5 mT (c), and for $|0\rangle\rightarrow|1\rangle$ transition at 2.3 mT (d) and 5 mT (e), each showing experimental data and light (dark) shaded area representing fitted Lorentzian to the $|0\rangle\rightarrow|-1\rangle$ ($|0\rangle\rightarrow|1\rangle$) component of the transition.}
\label{fig:figure4}
\end{figure}

We investigate the dependence of spin-state selectivity in V$_\text{B}^-$ defects on magnetic field strength. By varying the applied magnetic field by varying the separation between the permanent magnet and the sample and sweeping the microwave phase difference $\Delta$ at each field strength, we characterized the maximum achievable spin-state selectivity for each transition. Average ODMR peak separation is used to characterize the magnitude of the magnetic field. Figure 4 presents the maximum selectivities as functions of magnetic field. To quantify, the selectivity is defined for each of the two transition for any ODMR spectrum by fitting Lorentzian functions to the ODMR peaks, as shown in Fig.~\ref{fig:figure4}(b-e), and calculating the ratio of the area (light or dark, for $|0\rangle \to |-1\rangle$ and for $|0\rangle \to |1\rangle$ respectively) of the Lorentzian compared to the total fitted area (dark and light combined). Lorentzian fits were obtained using Origin software's double Lorentzian fitting routine applied on the spectra. The error bars represent the uncertainty in area calculations derived from the Lorentzian fitting. We also plot Lindblad calculation results (solid lines corresponding to the experimental data in dashed lines) in Figure 4 (a).

Lindblad calculations do not exhibit differences in maximum selectivities between the $|0\rangle \to |1\rangle$ and $|0\rangle \to |-1\rangle$ transitions. However, experimental results reveal distinctly different behavior for the $|0\rangle \to |-1\rangle$ transition compared to the $|0\rangle \to |1\rangle$ transition. Maximum spin-state selectivity is generally robust across the examined magnetic field range for the $|0\rangle \to |-1\rangle$ transition, with the highest selectivity achieved reaching $91.4\pm1.7\,\%$ at 1.1 mT. It is generally insensitive to the magnetic field below 5 mT while decreasing to approximately 60\% at the higher magnetic fields measured.

For both Lindblad calculations and experimental data for the maximum selectivity of $|0\rangle \to |1\rangle$ transition, maximum selectivity increases with magnetic field and eventually tends to plateau. The experimental maximum selectivity of $|0\rangle \to |1\rangle$ is relatively poor at low magnetic field, but ultimately reaches $83.5\pm1.9\,\%$ at 5 mT. This behavior in V$_\text{B}^-$ defects contrasts with that observed in NV$^-$ centers in diamond subjected to circularly polarized microwave. There, selectivity decreases with increasing magnetic field due to magnetic coupling between NV centers with different crystallographic orientations\cite{mrozek2015circularly}. In the case of spin defects which all share the same crystallographic orientation in V$_\text{B}^-$ hBN, higher magnetic field strengths reduce mixing between $|0\rangle \to |-1\rangle$ and $|0\rangle \to |1\rangle$ transitions, as demonstrated in the Fig. ~\ref{fig:figure4}(d-e). At 5 mT (Fig. ~\ref{fig:figure4}(e)), the ODMR signals show lessened overlap between transitions compared to the 2.3 mT case (Fig. ~\ref{fig:figure4}(d)), which exhibits greater overlap and consequently lower selectivity.

However, the magnetic field (in)dependence and the robustness of selectivity for the $|0\rangle \to |-1\rangle$ transition across varied magnetic fields is not captured in our numerical model. Possible sources of this discrepancy may include in-plane electric fields entering into the Hamiltonian as $d_{\perp}$ or $d^{\prime}$ terms \cite{candido2024interplay}, or a small difference in the intersystem crossing rates from $|e, \pm 1\rangle \rightarrow |s\rangle$ or $|s\rangle\rightarrow|g, \pm 1\rangle$. Elucidating this discrepancy is the focus of future work.

In conclusion, we have demonstrated spin-state selective excitation in V$_\text{B}^-$ defects in hBN using circularly polarized microwaves generated by a cross-shaped broadband waveguide. This approach expands the usefulness of the V$_\text{B}^-$ hBN platform for low magnetic field sensing and quantum information applications. Our experimental results, largely supported by Lindblad calculations, show continuous modulation of spin-state selectivity as a function of the phase difference between orthogonal microwave fields. Notably, an asymmetry in the maximum achievable selectivity was observed between the $|0\rangle \to |-1\rangle$ and $|0\rangle \to |1\rangle$ transitions, with the former exhibiting robustness across the tested magnetic field range and the latter showing improved selectivity at higher magnetic fields. The improved selectivity for the higher energy $|0\rangle \to |1\rangle$ transition at higher magnetic fields can be attributed to reduced mixing between the two transitions, as demonstrated by the lessened overlap in the ODMR signals. This behavior contrasts with that observed in NV$^-$ centers in diamond, where selectivity decreases with increasing magnetic field due to interactions between differently oriented defects. These findings demonstrate the potential of V$_\text{B}^-$ defects in hBN for applications requiring spin-state selectivity. Further investigations are necessary to fully understand the underlying physics of the selectivity behaviors of V$_\text{B}^-$ not captured by the Lindblad model, such as the asymmetry observed between the selectivities of the two transitions and the robustness of selectivity for the $|0\rangle \to |-1\rangle$ over a range of magnetic fields.

Supporting information containing detailed experimental methods including sample preparation, waveguide fabrication, and optical and microwave instrumentations, along with Lindblad model implementation details and analyses of magnetic field-dependent phase differences between maximally spin-state selective transitions in V$_\text{B}^-$ hBN defects is available.

We thank U.S. Department of Energy, Office of Science, National Quantum Information Science Research Centers, Quantum Science Center for experimental work at Purdue University. This work was also funded, in part, by the Laboratory Directed Research and Development Program and performed, in part, at the Center for Integrated Nanotechnologies, an Office of Science User Facility operated for the U.S. Department of Energy (DOE) Office of Science. Sandia National Laboratories is a multi-mission laboratory managed and operated by National Technology and Engineering Solutions of Sandia, LLC, a wholly owned subsidiary of Honeywell International, Inc., for the DOE’s National Nuclear Security Administration under contract DE-NA0003525. The calculation results are based upon work supported by the Air Force Office of Scientific Research under award number FA9550-22-1-0308. This paper describes objective technical results and analysis. Any subjective views or opinions that might be expressed in the paper do not necessarily represent the views of the U.S. Department of Energy or the United States Government. We also thank  Dr. Aroop Behera for assistance with 2D material transfer logistics, and Dr. Pauli Kehayias for early discussions on methods of generating circularly polarized microwave.

\bibliography{achemso-demo}

\providecommand{\latin}[1]{#1}
\makeatletter
\providecommand{\doi}
  {\begingroup\let\do\@makeother\dospecials
  \catcode`\{=1 \catcode`\}=2 \doi@aux}
\providecommand{\doi@aux}[1]{\endgroup\texttt{#1}}
\makeatother
\providecommand*\mcitethebibliography{\thebibliography}
\csname @ifundefined\endcsname{endmcitethebibliography}  {\let\endmcitethebibliography\endthebibliography}{}
\begin{mcitethebibliography}{3}
\providecommand*\natexlab[1]{#1}
\providecommand*\mciteSetBstSublistMode[1]{}
\providecommand*\mciteSetBstMaxWidthForm[2]{}
\providecommand*\mciteBstWouldAddEndPuncttrue
  {\def\EndOfBibitem{\unskip.}}
\providecommand*\mciteBstWouldAddEndPunctfalse
  {\let\EndOfBibitem\relax}
\providecommand*\mciteSetBstMidEndSepPunct[3]{}
\providecommand*\mciteSetBstSublistLabelBeginEnd[3]{}
\providecommand*\EndOfBibitem{}
\mciteSetBstSublistMode{f}
\mciteSetBstMaxWidthForm{subitem}{(\alph{mcitesubitemcount})}
\mciteSetBstSublistLabelBeginEnd
  {\mcitemaxwidthsubitemform\space}
  {\relax}
  {\relax}

\bibitem[Riendeau \latin{et~al.}(2023)Riendeau, Basso, Mah, Cong, Sadi, Henshaw, Azizur-Rahman, Jones, Joshi, Lilly, and Mounce]{riendeau2023quantum}
Riendeau,~E.~G.; Basso,~L.; Mah,~J.~J.; Cong,~R.; Sadi,~M.; Henshaw,~J.; Azizur-Rahman,~K.; Jones,~A.; Joshi,~G.; Lilly,~M.~P.; Mounce,~A.~A. Quantum Instrumentation Control Kit -- Defect Arbitrary Waveform Generator (QICK-DAWG): A Quantum Sensing Control Framework for Quantum Defects. 2023; \url{https://arxiv.org/abs/2311.18253}\relax
\mciteBstWouldAddEndPuncttrue
\mciteSetBstMidEndSepPunct{\mcitedefaultmidpunct}
{\mcitedefaultendpunct}{\mcitedefaultseppunct}\relax
\EndOfBibitem
\bibitem[Whitefield \latin{et~al.}(2024)Whitefield, Toth, Aharonovich, Tetienne, and Kianinia]{whitefield2024magnetic}
Whitefield,~B.; Toth,~M.; Aharonovich,~I.; Tetienne,~J.-P.; Kianinia,~M. Magnetic Field Sensitivity Optimization of Negatively Charged Boron Vacancy Defects in hBN. \emph{Advanced Quantum Technologies} \textbf{2024}, 2300118\relax
\mciteBstWouldAddEndPuncttrue
\mciteSetBstMidEndSepPunct{\mcitedefaultmidpunct}
{\mcitedefaultendpunct}{\mcitedefaultseppunct}\relax
\EndOfBibitem
\end{mcitethebibliography}


\providecommand{\latin}[1]{#1}
\makeatletter
\providecommand{\doi}
  {\begingroup\let\do\@makeother\dospecials
  \catcode`\{=1 \catcode`\}=2 \doi@aux}
\providecommand{\doi@aux}[1]{\endgroup\texttt{#1}}
\makeatother
\providecommand*\mcitethebibliography{\thebibliography}
\csname @ifundefined\endcsname{endmcitethebibliography}  {\let\endmcitethebibliography\endthebibliography}{}
\begin{mcitethebibliography}{36}
\providecommand*\natexlab[1]{#1}
\providecommand*\mciteSetBstSublistMode[1]{}
\providecommand*\mciteSetBstMaxWidthForm[2]{}
\providecommand*\mciteBstWouldAddEndPuncttrue
  {\def\EndOfBibitem{\unskip.}}
\providecommand*\mciteBstWouldAddEndPunctfalse
  {\let\EndOfBibitem\relax}
\providecommand*\mciteSetBstMidEndSepPunct[3]{}
\providecommand*\mciteSetBstSublistLabelBeginEnd[3]{}
\providecommand*\EndOfBibitem{}
\mciteSetBstSublistMode{f}
\mciteSetBstMaxWidthForm{subitem}{(\alph{mcitesubitemcount})}
\mciteSetBstSublistLabelBeginEnd
  {\mcitemaxwidthsubitemform\space}
  {\relax}
  {\relax}

\bibitem[Pirandola \latin{et~al.}(2018)Pirandola, Bardhan, Gehring, Weedbrook, and Lloyd]{pirandola2018advances}
Pirandola,~S.; Bardhan,~B.~R.; Gehring,~T.; Weedbrook,~C.; Lloyd,~S. Advances in photonic quantum sensing. \emph{Nature Photonics} \textbf{2018}, \emph{12}, 724--733\relax
\mciteBstWouldAddEndPuncttrue
\mciteSetBstMidEndSepPunct{\mcitedefaultmidpunct}
{\mcitedefaultendpunct}{\mcitedefaultseppunct}\relax
\EndOfBibitem
\bibitem[Schirhagl \latin{et~al.}(2014)Schirhagl, Chang, Loretz, and Degen]{schirhagl2014nitrogen}
Schirhagl,~R.; Chang,~K.; Loretz,~M.; Degen,~C.~L. Nitrogen-Vacancy Centers in Diamond: Nanoscale Sensors for Physics and Biology. \emph{Annual Review of Physical Chemistry} \textbf{2014}, \emph{65}, 83--105\relax
\mciteBstWouldAddEndPuncttrue
\mciteSetBstMidEndSepPunct{\mcitedefaultmidpunct}
{\mcitedefaultendpunct}{\mcitedefaultseppunct}\relax
\EndOfBibitem
\bibitem[Degen \latin{et~al.}(2017)Degen, Reinhard, and Cappellaro]{degen2017quantum}
Degen,~C.~L.; Reinhard,~F.; Cappellaro,~P. Quantum sensing. \emph{Reviews of Modern Physics} \textbf{2017}, \emph{89}, 035002\relax
\mciteBstWouldAddEndPuncttrue
\mciteSetBstMidEndSepPunct{\mcitedefaultmidpunct}
{\mcitedefaultendpunct}{\mcitedefaultseppunct}\relax
\EndOfBibitem
\bibitem[Budker and Romalis(2007)Budker, and Romalis]{budker2007optical}
Budker,~D.; Romalis,~M. Optical magnetometry. \emph{Nature Physics} \textbf{2007}, \emph{3}, 227--234\relax
\mciteBstWouldAddEndPuncttrue
\mciteSetBstMidEndSepPunct{\mcitedefaultmidpunct}
{\mcitedefaultendpunct}{\mcitedefaultseppunct}\relax
\EndOfBibitem
\bibitem[Romach \latin{et~al.}(2015)Romach, \latin{et~al.} others]{romach2015spectroscopy}
Romach,~Y.; others Spectroscopy of Surface-Induced Noise Using Shallow Spins in Diamond. \emph{Physical Review Letters} \textbf{2015}, \emph{114}, 017601\relax
\mciteBstWouldAddEndPuncttrue
\mciteSetBstMidEndSepPunct{\mcitedefaultmidpunct}
{\mcitedefaultendpunct}{\mcitedefaultseppunct}\relax
\EndOfBibitem
\bibitem[Sangtawesin \latin{et~al.}(2019)Sangtawesin, Dwyer, Srinivasan, Allred, Rodgers, De~Greve, Stacey, Dontschuk, O'Donnell, Hu, \latin{et~al.} others]{sangtawesin2019origins}
Sangtawesin,~S.; Dwyer,~B.~L.; Srinivasan,~S.; Allred,~J.~J.; Rodgers,~L.~V.; De~Greve,~K.; Stacey,~A.; Dontschuk,~N.; O'Donnell,~K.~M.; Hu,~D.; others Origins of Diamond Surface Noise Probed by Correlating Single-Spin Measurements with Surface Spectroscopy. \emph{Physical Review X} \textbf{2019}, \emph{9}, 031052\relax
\mciteBstWouldAddEndPuncttrue
\mciteSetBstMidEndSepPunct{\mcitedefaultmidpunct}
{\mcitedefaultendpunct}{\mcitedefaultseppunct}\relax
\EndOfBibitem
\bibitem[Azzam \latin{et~al.}(2021)Azzam, Parto, and Moody]{azzam2021prospects}
Azzam,~S.~I.; Parto,~K.; Moody,~G. Prospects and challenges of quantum emitters in 2D materials. \emph{Applied Physics Letters} \textbf{2021}, \emph{118}, 240502\relax
\mciteBstWouldAddEndPuncttrue
\mciteSetBstMidEndSepPunct{\mcitedefaultmidpunct}
{\mcitedefaultendpunct}{\mcitedefaultseppunct}\relax
\EndOfBibitem
\bibitem[Ren \latin{et~al.}(2019)Ren, Tan, and Zhang]{ren2019review}
Ren,~S.; Tan,~Q.; Zhang,~J. Review on the quantum emitters in two-dimensional materials. \emph{Journal of Semiconductors} \textbf{2019}, \emph{40}, 071903\relax
\mciteBstWouldAddEndPuncttrue
\mciteSetBstMidEndSepPunct{\mcitedefaultmidpunct}
{\mcitedefaultendpunct}{\mcitedefaultseppunct}\relax
\EndOfBibitem
\bibitem[Caldwell \latin{et~al.}(2019)Caldwell, Aharonovich, Cassabois, Edgar, Gil, and Basov]{caldwell2019photonics}
Caldwell,~J.~D.; Aharonovich,~I.; Cassabois,~G.; Edgar,~J.~H.; Gil,~B.; Basov,~D.~N. Photonics with hexagonal boron nitride. \emph{Nature Reviews Materials} \textbf{2019}, \emph{4}, 552--567\relax
\mciteBstWouldAddEndPuncttrue
\mciteSetBstMidEndSepPunct{\mcitedefaultmidpunct}
{\mcitedefaultendpunct}{\mcitedefaultseppunct}\relax
\EndOfBibitem
\bibitem[Kubanek(2022)]{kubanek2022coherent}
Kubanek,~A. Coherent Quantum Emitters in Hexagonal Boron Nitride. \emph{Advanced Quantum Technologies} \textbf{2022}, \emph{5}, 2200009\relax
\mciteBstWouldAddEndPuncttrue
\mciteSetBstMidEndSepPunct{\mcitedefaultmidpunct}
{\mcitedefaultendpunct}{\mcitedefaultseppunct}\relax
\EndOfBibitem
\bibitem[Cassabois \latin{et~al.}(2016)Cassabois, Valvin, and Gil]{cassabois2016hexagonal}
Cassabois,~G.; Valvin,~P.; Gil,~B. Hexagonal boron nitride is an indirect bandgap semiconductor. \emph{Nature Photonics} \textbf{2016}, \emph{10}, 262--266\relax
\mciteBstWouldAddEndPuncttrue
\mciteSetBstMidEndSepPunct{\mcitedefaultmidpunct}
{\mcitedefaultendpunct}{\mcitedefaultseppunct}\relax
\EndOfBibitem
\bibitem[Kianinia \latin{et~al.}(2017)Kianinia, \latin{et~al.} others]{kianinia2017robust}
Kianinia,~M.; others Robust Solid State Quantum System Operating at 800 K. Conference on Lasers and Electro-Optics. 2017; p JTu5A.24\relax
\mciteBstWouldAddEndPuncttrue
\mciteSetBstMidEndSepPunct{\mcitedefaultmidpunct}
{\mcitedefaultendpunct}{\mcitedefaultseppunct}\relax
\EndOfBibitem
\bibitem[Jungwirth \latin{et~al.}(2016)Jungwirth, Calderon, Ji, Spencer, Flatt{\'e}, and Fuchs]{jungwirth2016temperature}
Jungwirth,~N.~R.; Calderon,~B.; Ji,~Y.; Spencer,~M.~G.; Flatt{\'e},~M.~E.; Fuchs,~G.~D. Temperature Dependence of Wavelength Selectable Zero-Phonon Emission from Single Defects in Hexagonal Boron Nitride. \emph{Nano Letters} \textbf{2016}, \emph{16}, 6052--6057\relax
\mciteBstWouldAddEndPuncttrue
\mciteSetBstMidEndSepPunct{\mcitedefaultmidpunct}
{\mcitedefaultendpunct}{\mcitedefaultseppunct}\relax
\EndOfBibitem
\bibitem[Xue \latin{et~al.}(2018)Xue, \latin{et~al.} others]{xue2018anomalous}
Xue,~Y.; others Anomalous Pressure Characteristics of Defects in Hexagonal Boron Nitride Flakes. \emph{ACS Nano} \textbf{2018}, \emph{12}, 7127--7133\relax
\mciteBstWouldAddEndPuncttrue
\mciteSetBstMidEndSepPunct{\mcitedefaultmidpunct}
{\mcitedefaultendpunct}{\mcitedefaultseppunct}\relax
\EndOfBibitem
\bibitem[Gottscholl \latin{et~al.}(2020)Gottscholl, \latin{et~al.} others]{gottscholl2020initialization}
Gottscholl,~A.; others Initialization and read-out of intrinsic spin defects in a van der Waals crystal at room temperature. \emph{Nature Materials} \textbf{2020}, \emph{19}, 540--545\relax
\mciteBstWouldAddEndPuncttrue
\mciteSetBstMidEndSepPunct{\mcitedefaultmidpunct}
{\mcitedefaultendpunct}{\mcitedefaultseppunct}\relax
\EndOfBibitem
\bibitem[Reimers \latin{et~al.}(2020)Reimers, \latin{et~al.} others]{reimers2020photoluminescence}
Reimers,~J.~R.; others Photoluminescence, photophysics, and photochemistry of the \textbackslash{}mathrm\{V\}\_\textbackslash{}mathrm\{B\}\textasciicircum{}\{-\} defect in hexagonal boron nitride. \emph{Physical Review B} \textbf{2020}, \emph{102}, 144105\relax
\mciteBstWouldAddEndPuncttrue
\mciteSetBstMidEndSepPunct{\mcitedefaultmidpunct}
{\mcitedefaultendpunct}{\mcitedefaultseppunct}\relax
\EndOfBibitem
\bibitem[Chen and Quek(2021)Chen, and Quek]{chen2021photophysical}
Chen,~Y.; Quek,~S.~Y. Photophysical Characteristics of Boron Vacancy-Derived Defect Centers in Hexagonal Boron Nitride. \emph{Journal of Physical Chemistry C} \textbf{2021}, \emph{125}, 21791--21802\relax
\mciteBstWouldAddEndPuncttrue
\mciteSetBstMidEndSepPunct{\mcitedefaultmidpunct}
{\mcitedefaultendpunct}{\mcitedefaultseppunct}\relax
\EndOfBibitem
\bibitem[Iv{\'a}dy \latin{et~al.}(2020)Iv{\'a}dy, \latin{et~al.} others]{ivady2020ab}
Iv{\'a}dy,~V.; others Ab initio theory of the negatively charged boron vacancy qubit in hexagonal boron nitride. \emph{npj Computational Materials} \textbf{2020}, \emph{6}, 41\relax
\mciteBstWouldAddEndPuncttrue
\mciteSetBstMidEndSepPunct{\mcitedefaultmidpunct}
{\mcitedefaultendpunct}{\mcitedefaultseppunct}\relax
\EndOfBibitem
\bibitem[Gao \latin{et~al.}(2021)Gao, \latin{et~al.} others]{gao2021high}
Gao,~X.; others High-Contrast Plasmonic-Enhanced Shallow Spin Defects in Hexagonal Boron Nitride for Quantum Sensing. \emph{Nano Letters} \textbf{2021}, \emph{21}, 7708--7714\relax
\mciteBstWouldAddEndPuncttrue
\mciteSetBstMidEndSepPunct{\mcitedefaultmidpunct}
{\mcitedefaultendpunct}{\mcitedefaultseppunct}\relax
\EndOfBibitem
\bibitem[Yu \latin{et~al.}(2022)Yu, \latin{et~al.} others]{yu2022excited}
Yu,~P.; others Excited-State Spectroscopy of Spin Defects in Hexagonal Boron Nitride. \emph{Nano Letters} \textbf{2022}, \emph{22}, 3545--3549\relax
\mciteBstWouldAddEndPuncttrue
\mciteSetBstMidEndSepPunct{\mcitedefaultmidpunct}
{\mcitedefaultendpunct}{\mcitedefaultseppunct}\relax
\EndOfBibitem
\bibitem[Mu \latin{et~al.}(2022)Mu, \latin{et~al.} others]{mu2022excited}
Mu,~Z.; others Excited-State Optically Detected Magnetic Resonance of Spin Defects in Hexagonal Boron Nitride. \emph{Physical Review Letters} \textbf{2022}, \emph{128}, 216402\relax
\mciteBstWouldAddEndPuncttrue
\mciteSetBstMidEndSepPunct{\mcitedefaultmidpunct}
{\mcitedefaultendpunct}{\mcitedefaultseppunct}\relax
\EndOfBibitem
\bibitem[Mathur \latin{et~al.}(2022)Mathur, \latin{et~al.} others]{mathur2022excited}
Mathur,~N.; others Excited-state spin-resonance spectroscopy of V$_{{\mathrm{B}}}$$^{-}$ defect centers in hexagonal boron nitride. \emph{Nature Communications} \textbf{2022}, \emph{13}, 3233\relax
\mciteBstWouldAddEndPuncttrue
\mciteSetBstMidEndSepPunct{\mcitedefaultmidpunct}
{\mcitedefaultendpunct}{\mcitedefaultseppunct}\relax
\EndOfBibitem
\bibitem[Nelson(1957)]{nelson1957circularly}
Nelson,~C.~E. Circularly Polarized Microwave Cavity Filters. \emph{IRE Transactions on Microwave Theory and Techniques} \textbf{1957}, \emph{5}, 136--147\relax
\mciteBstWouldAddEndPuncttrue
\mciteSetBstMidEndSepPunct{\mcitedefaultmidpunct}
{\mcitedefaultendpunct}{\mcitedefaultseppunct}\relax
\EndOfBibitem
\bibitem[Henderson \latin{et~al.}(2008)Henderson, Ramsey, Quddusi, and del Barco]{henderson2008high}
Henderson,~J.~J.; Ramsey,~C.~M.; Quddusi,~H.~M.; del Barco,~E. High-frequency microstrip cross resonators for circular polarization electron paramagnetic resonance spectroscopy. \emph{Review of Scientific Instruments} \textbf{2008}, \emph{79}, 074704\relax
\mciteBstWouldAddEndPuncttrue
\mciteSetBstMidEndSepPunct{\mcitedefaultmidpunct}
{\mcitedefaultendpunct}{\mcitedefaultseppunct}\relax
\EndOfBibitem
\bibitem[Alegre \latin{et~al.}(2007)Alegre, Santori, Medeiros-Ribeiro, and Beausoleil]{alegre2007polarization}
Alegre,~T. P.~M.; Santori,~C.; Medeiros-Ribeiro,~G.; Beausoleil,~R.~G. Polarization-selective excitation of nitrogen vacancy centers in diamond. \emph{Physical Review B} \textbf{2007}, \emph{76}, 165205\relax
\mciteBstWouldAddEndPuncttrue
\mciteSetBstMidEndSepPunct{\mcitedefaultmidpunct}
{\mcitedefaultendpunct}{\mcitedefaultseppunct}\relax
\EndOfBibitem
\bibitem[Yaroshenko \latin{et~al.}(2020)Yaroshenko, \latin{et~al.} others]{yaroshenko2020circularly}
Yaroshenko,~V.; others Circularly polarized microwave antenna for nitrogen vacancy centers in diamond. \emph{Review of Scientific Instruments} \textbf{2020}, \emph{91}, 035003\relax
\mciteBstWouldAddEndPuncttrue
\mciteSetBstMidEndSepPunct{\mcitedefaultmidpunct}
{\mcitedefaultendpunct}{\mcitedefaultseppunct}\relax
\EndOfBibitem
\bibitem[Riendeau \latin{et~al.}(2023)Riendeau, Basso, Mah, Cong, Sadi, Henshaw, Azizur-Rahman, Jones, Joshi, Lilly, and Mounce]{riendeau2023quantum}
Riendeau,~E.~G.; Basso,~L.; Mah,~J.~J.; Cong,~R.; Sadi,~M.; Henshaw,~J.; Azizur-Rahman,~K.; Jones,~A.; Joshi,~G.; Lilly,~M.~P.; Mounce,~A.~A. Quantum Instrumentation Control Kit -- Defect Arbitrary Waveform Generator (QICK-DAWG): A Quantum Sensing Control Framework for Quantum Defects. 2023; \url{https://arxiv.org/abs/2311.18253}\relax
\mciteBstWouldAddEndPuncttrue
\mciteSetBstMidEndSepPunct{\mcitedefaultmidpunct}
{\mcitedefaultendpunct}{\mcitedefaultseppunct}\relax
\EndOfBibitem
\bibitem[Gao \latin{et~al.}(2022)Gao, Vaidya, Li, Ju, Jiang, Xu, Llacsahuanga~Allcca, Shen, Taniguchi, Watanabe, Bhave, Chen, Ping, and Li]{gao2022nuclear}
Gao,~X.; Vaidya,~S.; Li,~K.; Ju,~P.; Jiang,~B.; Xu,~Z.; Llacsahuanga~Allcca,~A.~E.; Shen,~K.; Taniguchi,~T.; Watanabe,~K.; Bhave,~S.~A.; Chen,~Y.~P.; Ping,~Y.; Li,~T. Nuclear spin polarization and control in hexagonal boron nitride. \emph{Nature Materials} \textbf{2022}, \emph{21}, 1024--1028\relax
\mciteBstWouldAddEndPuncttrue
\mciteSetBstMidEndSepPunct{\mcitedefaultmidpunct}
{\mcitedefaultendpunct}{\mcitedefaultseppunct}\relax
\EndOfBibitem
\bibitem[Gracheva \latin{et~al.}(2023)Gracheva, \latin{et~al.} others]{gracheva2023symmetry}
Gracheva,~I.~N.; others Symmetry of the Hyperfine and Quadrupole Interactions of Boron Vacancies in a Hexagonal Boron Nitride. \emph{Journal of Physical Chemistry C} \textbf{2023}, \emph{127}, 3634--3639\relax
\mciteBstWouldAddEndPuncttrue
\mciteSetBstMidEndSepPunct{\mcitedefaultmidpunct}
{\mcitedefaultendpunct}{\mcitedefaultseppunct}\relax
\EndOfBibitem
\bibitem[Candido and Flatt\'e(2024)Candido, and Flatt\'e]{candido2024theory}
Candido,~D.~R.; Flatt\'e,~M.~E. Theory of spin center sensing of diffusion and other surface electric dynamics. \emph{Phys. Rev. B} \textbf{2024}, \emph{110}, 174450\relax
\mciteBstWouldAddEndPuncttrue
\mciteSetBstMidEndSepPunct{\mcitedefaultmidpunct}
{\mcitedefaultendpunct}{\mcitedefaultseppunct}\relax
\EndOfBibitem
\bibitem[Patel \latin{et~al.}(2024)Patel, Fishman, Huang, Gusdorff, Fehr, Hopper, Breitweiser, Porat, Flatté, and Bassett]{patel2024room}
Patel,~R.~N.; Fishman,~R. E.~K.; Huang,~T.-Y.; Gusdorff,~J.~A.; Fehr,~D.~A.; Hopper,~D.~A.; Breitweiser,~S.~A.; Porat,~B.; Flatté,~M.~E.; Bassett,~L.~C. Room Temperature Dynamics of an Optically Addressable Single Spin in Hexagonal Boron Nitride. \emph{Nano Letters} \textbf{2024}, \emph{24}, 7623--7628\relax
\mciteBstWouldAddEndPuncttrue
\mciteSetBstMidEndSepPunct{\mcitedefaultmidpunct}
{\mcitedefaultendpunct}{\mcitedefaultseppunct}\relax
\EndOfBibitem
\bibitem[Candido and Flatté(2024)Candido, and Flatté]{candido2024interplay}
Candido,~D.~R.; Flatté,~M.~E. Interplay between charge and spin noise in the near-surface theory of decoherence and relaxation of ${C}_{3v}$ symmetry qutrit spin-1 centers. \emph{Phys. Rev. B} \textbf{2024}, \emph{110}, 024419\relax
\mciteBstWouldAddEndPuncttrue
\mciteSetBstMidEndSepPunct{\mcitedefaultmidpunct}
{\mcitedefaultendpunct}{\mcitedefaultseppunct}\relax
\EndOfBibitem
\bibitem[Elko \latin{et~al.}(2024)Elko, Hassenmayer, Higgins, Lenahan, Flatté, Fehr, Craven, and Larsen]{elko2024near}
Elko,~M.~J.; Hassenmayer,~D.~T.; Higgins,~A.~A.; Lenahan,~P.~M.; Flatté,~M.~E.; Fehr,~D.; Craven,~M.~D.; Larsen,~T.~D. Near zero-field magnetoresistance and defects in gallium nitride pn junctions. \emph{Journal of Vacuum Science \& Technology B} \textbf{2024}, \emph{42}, 052205\relax
\mciteBstWouldAddEndPuncttrue
\mciteSetBstMidEndSepPunct{\mcitedefaultmidpunct}
{\mcitedefaultendpunct}{\mcitedefaultseppunct}\relax
\EndOfBibitem
\bibitem[Whitefield \latin{et~al.}(2024)Whitefield, Toth, Aharonovich, Tetienne, and Kianinia]{whitefield2024magnetic}
Whitefield,~B.; Toth,~M.; Aharonovich,~I.; Tetienne,~J.-P.; Kianinia,~M. Magnetic Field Sensitivity Optimization of Negatively Charged Boron Vacancy Defects in hBN. \emph{Advanced Quantum Technologies} \textbf{2024}, 2300118\relax
\mciteBstWouldAddEndPuncttrue
\mciteSetBstMidEndSepPunct{\mcitedefaultmidpunct}
{\mcitedefaultendpunct}{\mcitedefaultseppunct}\relax
\EndOfBibitem
\bibitem[Mrózek \latin{et~al.}(2015)Mrózek, Mlynarczyk, Rudnicki, and Gawlik]{mrozek2015circularly}
Mrózek,~M.; Mlynarczyk,~J.; Rudnicki,~D.~S.; Gawlik,~W. Circularly polarized microwaves for magnetic resonance study in the GHz range: Application to nitrogen-vacancy in diamonds. \emph{Applied Physics Letters} \textbf{2015}, \emph{107}, 013505\relax
\mciteBstWouldAddEndPuncttrue
\mciteSetBstMidEndSepPunct{\mcitedefaultmidpunct}
{\mcitedefaultendpunct}{\mcitedefaultseppunct}\relax
\EndOfBibitem
\end{mcitethebibliography}

\end{document}


\section{Experimental setup}

\subsection{Sample Preparation}
Commercial hBN flakes (2D Semiconductors) were exfoliated onto a silicon substrate and irradiated with 2.5 keV He+ ions at a fluence of 0.2 nm$^{-2}$ using an ISIS3000 ion implanter (PSP Vacuum Technology) to generate (V$_\text{B}^-$) defects. The irradiated hBN flakes were then transferred onto a gold waveguide using a polycarbonate (PC) stamp mounted on a polydimethylsiloxane (PDMS) block. The transfer process involved heating the substrate to 160 °C, cooling to 50 °C, and lifting the PC stamp with the hBN flakes. The stamp was then brought into contact with the gold substrate, heated to 190 °C to melt the PC film, and the PDMS was removed. The substrate was cleaned with chloroform, acetone, and isopropanol to ensure clean deposition of the hBN flakes.
\subsection{Waveguide Fabrication}
The gold on sapphire cross-shaped waveguide was fabricated using photolithographic patterning with an MLA 150 Maskless Aligner, followed by gold deposition using a metal evaporator. The printed circuit board (PCB) for connecting the waveguide was fabricated using an LPKF Protolaser U4.
\subsection{Optical and microwave signal generation and processing}
The optical configuration for capturing the ODMR signal (Fig. S1) consists of a green laser of 532 nm for sample illumination, and longpass filter for red light detection coupled to a single photon counter (SPCM) for V$_\text{B}^-$ hBN experiments. Microwave signal generation, amplification, and transmission, as well as data collection and processing, are performed by the FPGA ZCU111 in combination with the Quantum Instrument Control Kit - Defect Arbitrary Waveform Generator (QICK-DAWG). QICK-DAWG\cite{riendeau2023quantum}, an open-source software package developed at Sandia National Laboratories, enables laser control, microwave pulsing, and low-frequency readout for NV control and measurement. The package is adopted and modified to be suitable for V$_\text{B}^-$ hBN experiments. A translation stage with a fixed permanent magnet is employed for the V$_\text{B}^-$ hBN experiments.
\setcounter{figure}{0}
\begin{figure}[H]
    \centering
    \renewcommand{\thefigure}{S\arabic{figure}}
    \includegraphics[width=0.95\textwidth]{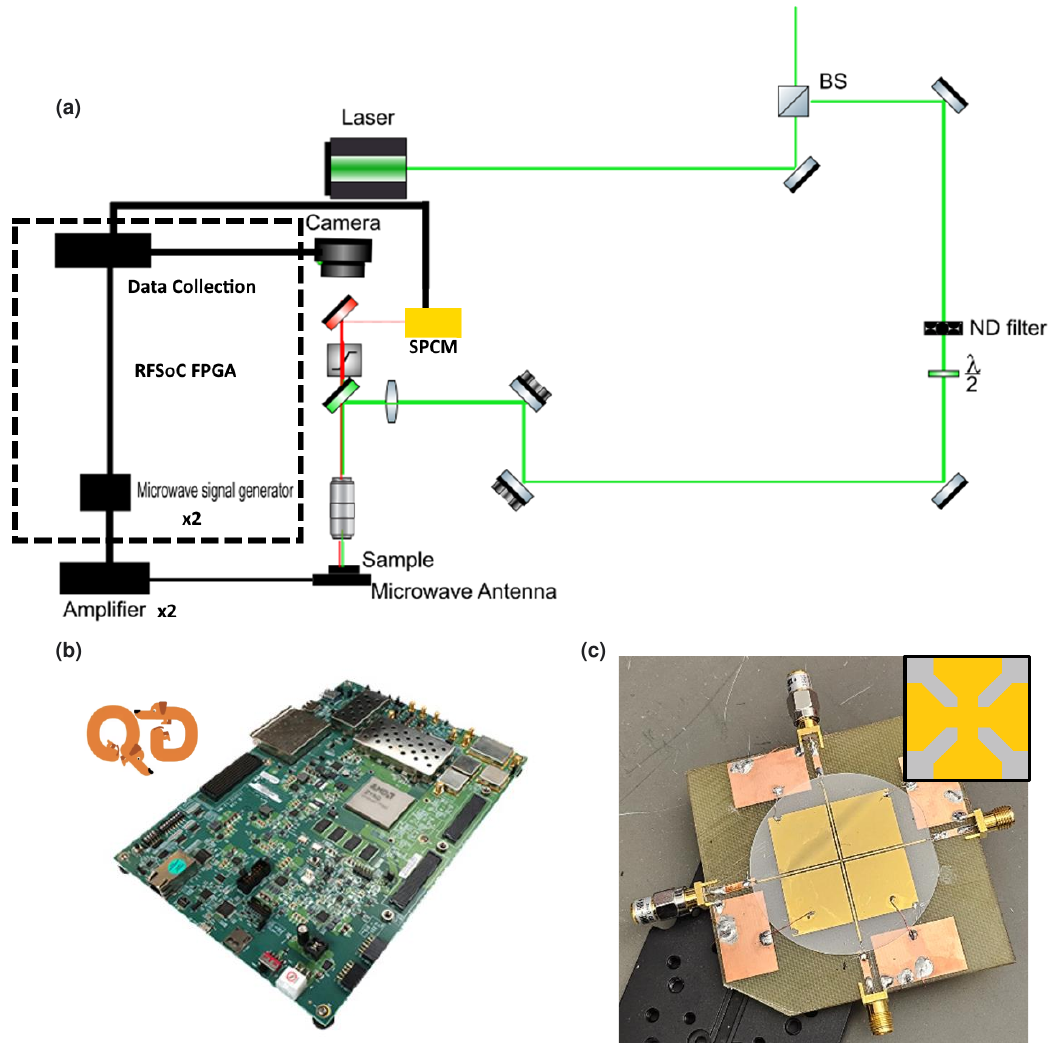}
    \caption{Experimental setup for phase-synced dual microwave excitation of V$_\text{B}^-$ defects in hBN. (a) Schematic diagram of the experimental apparatus, including optical components, microwave generation and delivery systems, and data processing units. (b) ZCU111 RFSoC FPGA controlled using the Qick-Dawg package for synchronized microwave generation, precise timing control, and real-time data processing. (c) Gold cross-waveguide is deposited on sapphire substrate, with $50~\Omega$ terminations at the ends of orthogonal microwave transmission lines. Rectangular gold pads serve as ground planes. The inset shows that the waveguide narrows to a $50~\mu\text{m}$ width at the center to enhance B$_\text{MW}$ delivery. The device is mounted on a laser-etched copper printed circuit board.}
    \label{fig:figureS1}
\end{figure}

\section{Lindblad Model Implementation}
Lindblad calculations were completed with a 7-level model and similar optical rates as Whitefield et al.\cite{whitefield2024magnetic}. In the notation of Whitefield et al. \cite{whitefield2024magnetic}, we used $k_{p}=7\ \mu s^{-1}$, $k_{d}=880\ \mu s^{-1}$, $k_{45}=1.15\ ns^{-1}$, $k_{35}=220\ \mu s^{-1}$, $k_{52}=20\ \mu s^{-1}$, and $k_{51}=13\ \mu s^{-1}$. In adapting the model and rates from Whitefield et al. \cite{whitefield2024magnetic} to the 7-level model used in this work, we assumed the following correspondence: $\langle 4|\hat{\rho}|4\rangle\simeq\langle e|\langle +1|\hat{\rho}|+1\rangle|e\rangle+\langle e|\langle -1|\hat{\rho}|-1\rangle|e\rangle$ and $\langle 2|\hat{\rho}|2\rangle\simeq\langle g|\langle +1|\hat{\rho}|+1\rangle|g\rangle+\langle g|\langle -1|\hat{\rho}|-1\rangle|g\rangle$. In other words, the population in state $|4\rangle$ is equivalent to the total population in states $|e\rangle|\pm1\rangle$, and the population in state $|2\rangle$ is equivalent to the total population in states $|g\rangle|\pm1\rangle$. Finally, we adjusted $k_{35}$ from its value in Whitefield et al. \cite{whitefield2024magnetic} to allow the calculated Contrast to match the measured Contrast in this work.

\section*{Magnetic field-dependent deviations in microwave phase difference between spin-state selective transitions in V$_\text{B}^-$ hBN}

\begin{figure}[H]
    \centering
    \renewcommand{\thefigure}{S\arabic{figure}}
    \includegraphics[width=0.95\textwidth]{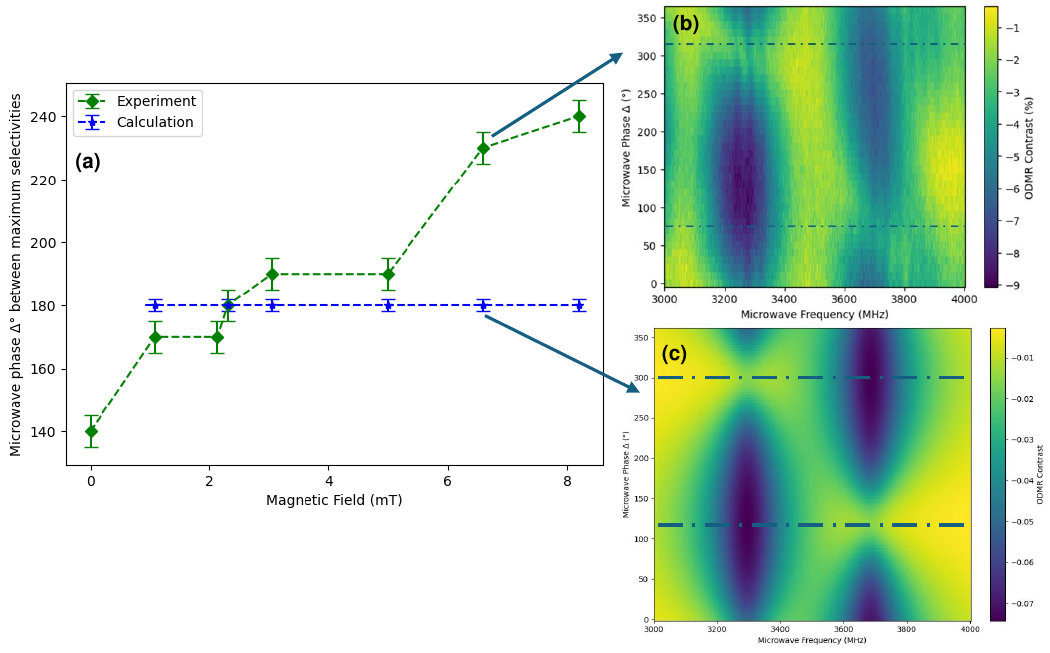}
    \caption{Magnetic field dependence of phase separation between maximum spin-state selectivities. (a) Relationship between observed phase differences ($\Delta^\circ$) of orthogonal linearly polarized microwaves for maximum spin-state selectivities and the applied magnetic field into the device, showing experimental data (green) with error bars from microwave phase step size and theoretical calculations (blue) that fail to capture the observed deviations from 180° at field extremes. (b) Color plot of experimental ODMR spectra evolution versus phase difference ($\Delta^\circ$) at 6.58 mT into the device, with blue dashed lines marking maximum selectivity conditions. (c) Corresponding Lindblad calculations at 6.58 mT showing 180$^\circ$ phase separation between maximum selectivity conditions, differing from experimental data in (b). 
}
    \label{fig:figureS2}
\end{figure}
Our measurements reveal unexpected variations in the phase differences between the two $\Delta$s each corresponding to maximum selectivity for the two spin-state transitions individually. These show significant deviations from the expected 180° separation predicted by Equation 3 in the main manuscript. For magnetic field pointing into the device, this 180° separation appears to hold approximately true only within a limited magnetic field range of 2-5 mT. At zero field, we observe a smaller phase separation of only 140°, while at higher fields reaching 8.2 mT, the separation increases significantly to 240°. The error bars shown in our experimental measurements reflect the microwave phase step size used during the scanning process.
The comparison between experimental ODMR spectra evolution and theoretical predictions at 6.58 mT highlights this discrepancy. While experimental data shows variable phase separation between maximum selectivity conditions marked by blue dashed lines, the corresponding Lindblad calculations show a constant phase separation that differs from experimental observations. In case of zero magnetic field, there is negligible change in selectivities of the two separate spin-state transitions with microwave phase change during calculation, with both transitions equally favored, so $\Delta$ can be stated as undefined.
\begin{figure}[H]
    \centering
    \renewcommand{\thefigure}{S\arabic{figure}}
    \includegraphics[width=0.95\textwidth]{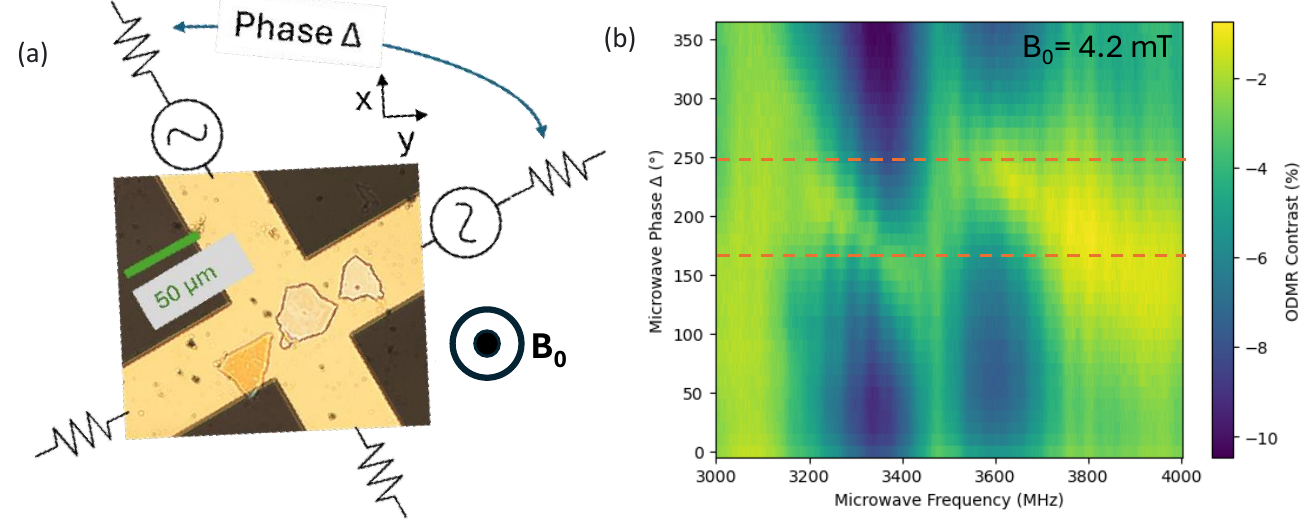}
    \caption{Inverted magnetic field orientation affecting spin-state selectivity in V$_\text{B}^-$ defects in hBN. (a) Experimental configuration with the magnetic field oriented coming out of the plane, in contrast to the going into-the-plane orientation described in the main manuscript. (b) Spin-state selectivity under this reversed field configuration, demonstrating around 90° phase difference ($\Delta^\circ$) separation between maximum selectivities (indicated by dashed oranged lines), as opposed to the approximately 180° separation observed in the main manuscript for the in-plane field orientation for the same magnetic field strength.
}
    \label{fig:figureS3}
\end{figure}
In our investigation of an inverted magnetic field configuration with the field oriented coming out of the plane, we observe maximum spin-state selectivities in V$_\text{B}^-$ hBN separated by a microwave phase difference ($\Delta$) of around 90°, which contrasts with the approximately 180° separation observed for similar strength into-the-plane fields.
These observed deviations in both standard and inverted field configurations might potentially be related to similar underlying factors. Changes in the effective angle between the defect symmetry axis and the microwave magnetic field B$_\text{MW}$ could play a role, as this angle may not remain perpendicular at all field strengths or orientations. This misalignment might result in the projection of supposedly circular microwave polarization in the x-y plane onto the defect axis as elliptical and vice versa. Our theoretical calculations do not capture these observed deviations, suggesting additional factors may be involved that are not accounted for in our current models. Various additional factors could include stray fields, distortions of defects, or device-induced alterations to the plane where B$\text{MW}$ is rotating, though these have not been fully characterized in this study. 
Our observations suggest that there may be a complex interplay between applied magnetic field, microwave polarization, and the atomic structure of V$_\text{B}^-$ defects in hBN. Further investigation into these phenomena could yield insights for improving device performance and understanding of the defects.

\bibliography{achemso-demo}